\documentclass[sigconf, nonacm]{acmart}

\newcommand\vldbdoi{XX.XX/XXX.XX}

\newcommand\vldbavailabilityurl{https://github.com/Dingyi-Kang/LAANN}

\usepackage{graphicx}
\usepackage{amsmath}

\usepackage{amssymb}
\usepackage{amsfonts}
\usepackage{booktabs}
\usepackage{multirow}
\usepackage{xcolor}
\usepackage{soul}
\usepackage{siunitx}
\usepackage{enumitem}
\usepackage{makecell}
\usepackage{subcaption}
\usepackage{algorithm}
\usepackage{algpseudocode}
\usepackage{threeparttable}
\usepackage{balance}
\usepackage{float}

\newcommand{\circled}[1]{\textcircled{\scriptsize #1}}

\begin{document}
\title{LAANN: I/O-Aware Look-Ahead Search for Disk-Based Approximate Nearest Neighbor Search}

\author{Dingyi Kang}
\affiliation{\institution{University of Texas at Dallas}}
\email{dingyi.kang@utdallas.edu}

\author{Juncheng Yang}
\affiliation{\institution{Harvard University}}
\email{juncheng@seas.harvard.edu}

\author{Bingzhe Li}
\affiliation{\institution{University of Texas at Dallas}}
\email{bingzhe.li@utdallas.edu}

\begin{abstract}

Approximate nearest neighbor search (ANNS) is a fundamental primitive in large-scale retrieval, recommendation, and AI systems. As vector datasets grow to billions or even trillions of items, disk-based ANNS systems have emerged to handle this scale by storing vector data and index structures on storage systems, but their query performance remains dominated by I/O latency. Existing disk-based ANNS systems primarily optimize I/O efficiency or overlap I/O with computation, but they treat CPU computation and I/O access as largely separate components. This separation misses a critical opportunity: selectively processing candidates already cached in memory before making I/O decisions can reduce unnecessary disk accesses and improve search quality. However, exploiting this opportunity is challenging because excessive computation can delay critical I/O operations, while poorly chosen computation provides little benefit, potentially increasing overall query latency.

In this paper, we present \emph{LAANN}, a disk-based ANNS system that makes graph search explicitly I/O-aware by co-optimizing CPU computation and I/O access. LAANN combines three techniques: \emph{look-ahead search}, which adapts the search strategy across query stages to balance I/O reduction and timely I/O issuance; a \emph{priority I/O--CPU pipeline}, which uses I/O waiting time to process candidates cached in memory according to their expected impact on upcoming I/O decisions; and a \emph{fast lightweight in-memory graph index}, which provides high-quality initial candidates to accelerate convergence and reduce disk accesses. Experiments on million- and billion-scale datasets demonstrate that LAANN substantially outperforms state-of-the-art disk-based ANNS systems. At Recall@10 = 0.9, LAANN achieves $1.41\times$--$4.66\times$ higher throughput, 29\%--79\% lower latency, and $1.59\times$--$6.34\times$ fewer I/O operations.
\end{abstract}

\maketitle



\renewcommand\thefootnote{}\footnotetext{Artifact availability: 
\url{https://github.com/Dingyi-Kang/PageANN}.}\renewcommand\thefootnote{\arabic{footnote}}

\begin{figure}[!t]
    \centering
    \includegraphics[width=0.999\linewidth]{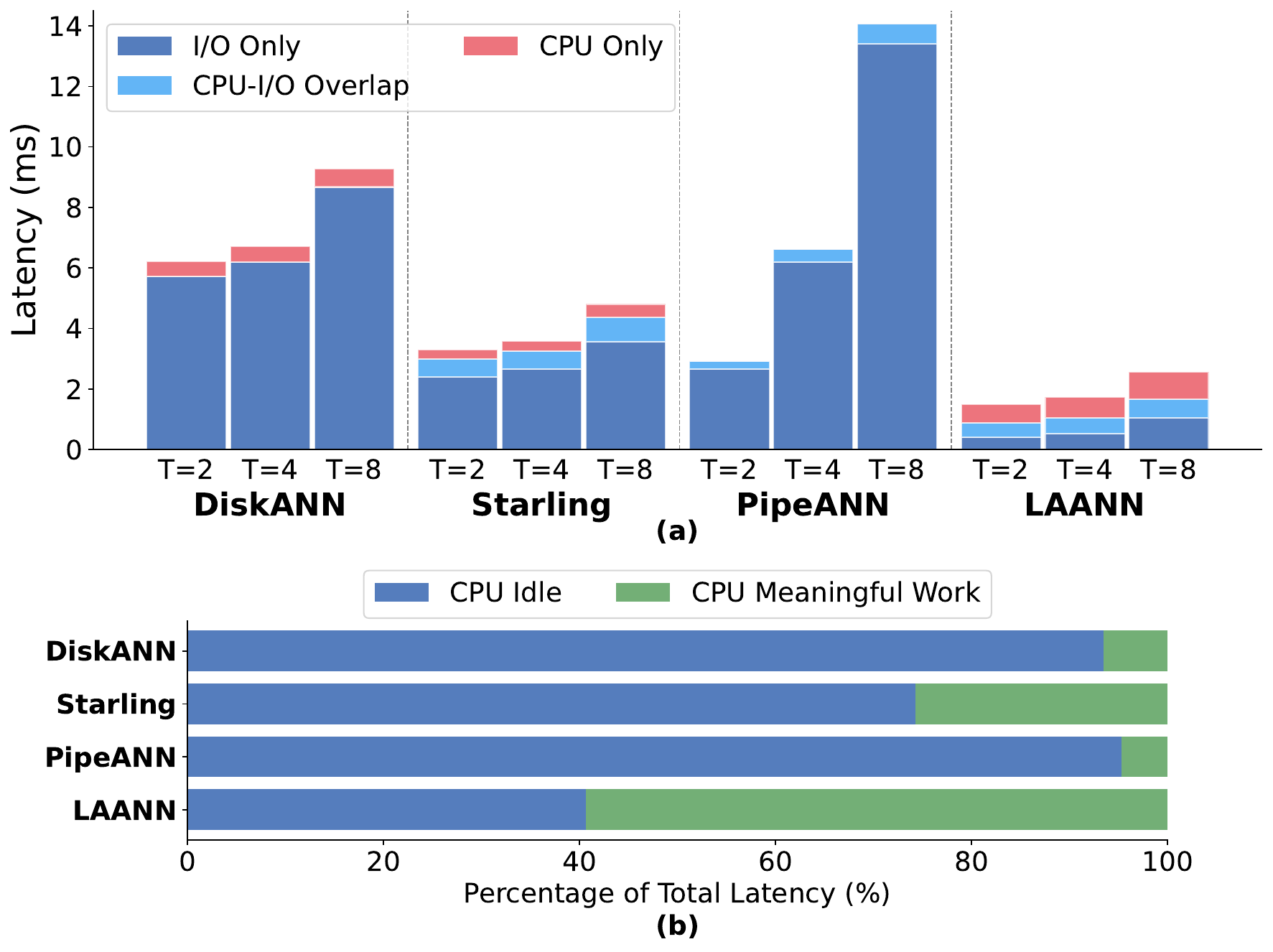}
    \caption{Latency breakdown on SIFT100M under a memory budget of $0.4\times$ dataset size with $T \in \{2,4,8\}$ threads. (a) Latency decomposed into I/O only, CPU-I/O overlap, and CPU only components. (b) Breakdown of total query latency into CPU meaningful work and CPU idle time at $T=8$ threads.}
    \label{fig:io_cpu_breakdown}
\end{figure}
\section{Introduction}\label{sec:introduction}


Approximate Nearest Neighbor Search (ANNS) finds the most similar vectors to a query from a dataset of millions to billions of points, serving as a core building block in machine learning~\cite{cao2017binary, bijalwan2014knn, tagami2017annexml, cover1967nearest, zhu2019accelerating, flickner1995query, zhang2022uni, zhang2018visual, huang2020embedding}, retrieval-augmented generation~\cite{lewis2020retrieval, guu2020retrieval, borgeaud2022improving, asai2023retrieval, liu2023learning, ram2023context, izacard2023atlas}, and bioinformatics~\cite{bittremieux2018fast, schutze2022nearest} systems. Among various ANNS approaches, graph-based methods~\cite{malkov2014approximate, hnsw16, chen2018sptag, fu2017fast, nsg19, li2019approximate} achieve the best tradeoff between search speed and recall, with state-of-the-art (SOTA) in-memory indexes such as HNSW~\cite{hnsw16} and Vamana~\cite{diskann} achieving high recall with sub-millisecond latency. However, as datasets scale to billions or trillions of vectors, these methods face significant scalability challenges, as both the dataset and the graph index grow proportionally with the number of vectors. Disk-based ANNS methods address this by offloading vectors and graph edges to SSD, reducing the memory requirement to a small fraction of the dataset size, but at the cost of high I/O latency.

Existing disk-based ANNS methods reduce I/O latency through a range of optimizations, including disk layout optimization~\cite{diskann++, starling, margo}, I/O scheduling~\cite{pipeann}, in-memory entry point optimization~\cite{starling, margo, pipeann}, and caching frequently visited nodes in memory~\cite{diskann, starling, margo, pageann}. However, as shown in Figure~\ref{fig:io_cpu_breakdown}(a), even when these optimizations are combined, I/O latency still accounts for over 90\% of total query latency. Furthermore, this I/O bottleneck worsens as the number of concurrent threads increases due to I/O contention, which is most severe for PipeANN as it intentionally issues more I/O operations to maximize I/O overlap, further intensifying contention and causing significant latency degradation at higher thread counts. Meanwhile, as shown in Figure~\ref{fig:io_cpu_breakdown}(b), the CPU remains idle for more than 74.3\% of query time even for the best-performing baseline. This indicates that existing systems leave substantial performance on the table by treating CPU computation and I/O access as largely separate concerns. They address I/O latency mainly from the I/O perspective, whether by reducing I/O read amplification or overlapping I/O operations. However, they overlook the orthogonal opportunity to use CPU computation, which is lightweight and largely underutilized, to reduce the total number of I/O operations.

One of our key observations is that CPU computation and I/O operations are tightly coupled: \textit{pre-processing vectors that are cached in memory before making subsequent I/O decisions can reduce the total number of disk accesses}. In particular, expanding candidates that are not currently ranked closest to the query can still advance the search frontier and help the search navigate toward the true nearest neighbors with fewer I/Os. Moreover, because disk-based graph traversal relies on approximate distance estimates, lower-ranked candidates may still turn out to be true nearest neighbors under exact distance computation. Processing these candidates can therefore improve search accuracy and reduce the number of I/O operations required to reach a target recall.

However, simply processing more cached vectors creates a fundamental tradeoff. On the one hand, additional CPU computation can reduce future disk accesses by improving search decisions. On the other hand, excessive computation increases CPU overhead and can delay the submission of new I/O requests, thereby increasing end-to-end query latency. Conversely, processing too few candidates leaves potential I/O reductions unexploited. The key challenge is therefore to balance CPU overhead, I/O reduction, and timely I/O submission during disk-based graph search.

A natural way to exploit this tradeoff is to shift additional CPU computation into I/O waiting periods, when the CPU is largely idle in existing systems. However, doing so effectively raises two practical challenges. First, cached candidates differ significantly in their usefulness. Some candidates may directly affect the next I/O decision and should be processed before issuing new disk requests; others can be postponed without hurting search progress; and some are unlikely to contribute to the search at all. Thus, the system must decide not only how much extra computation to perform, but also which candidates should receive priority. Second, the number of useful candidates cached in the memory available during each I/O wait is limited. Once these candidates are processed, the CPU can quickly become idle again, leaving much of the I/O waiting time unexploited.

To address these challenges, we present \emph{LAANN}, a disk-based ANNS system that makes graph search explicitly I/O-aware. The look-ahead search resolves the dilemma of balancing I/O reduction with timely I/O submission by adapting its strategy to the current search phase. The priority I/O-CPU pipeline addresses the challenge of limited useful CPU work during I/O waiting by systematically filling idle CPU time with work ordered by relevance to the next I/O decision, supported by a customized candidate pool with an overflow area. Additionally, we observe that existing in-memory graph indexes suffer from a precision mismatch with disk-based graph search, forcing the search to start from a nearly empty candidate pool and wasting I/Os in the approach phase. To address this, we propose a \emph{fast lightweight in-memory graph index} that seeds the disk graph search with high-quality candidates from the start, reducing approach phase I/Os. Together, these three techniques address the I/O bottleneck by exploiting the coupling between CPU computation and I/O operations.

\noindent This paper makes the following contributions:
\begin{itemize}[leftmargin=*, itemsep=2pt, topsep=4pt]
\item We identify the tight coupling between CPU computation and I/O operations and analyze the fundamentally different I/O characteristics of the approach and convergence phases.
\item We propose look-ahead search, a priority I/O-CPU pipeline with overflow candidate pool, and a fast lightweight in-memory graph index.
\item We evaluate LAANN on five benchmark datasets at 100M and 1B scale against five SOTA baselines, demonstrating $1.41\times$--$4.66\times$ higher throughput, 29\%--79\% lower latency, and $1.59\times$--$6.34\times$ fewer I/O operations.
\end{itemize}

The rest of this paper is organized as follows. Section~\ref{sec:background} provides background on graph-based and disk-based ANNS. Section~\ref{sec:motivation} analyzes the I/O bottleneck, identifies the coupling between CPU work and I/O operations, and characterizes the challenges of exploiting this coupling. Section~\ref{sec:design} presents the design of LAANN. Section~\ref{sec:implementation} describes the implementation. Section~\ref{sec:results} evaluates LAANN against five SOTA disk-based ANNS baselines. Section~\ref{sec:related} discusses related work, and Section~\ref{sec:conclusion} concludes.


\section{Background}\label{sec:background}
\subsection{Graph-based ANNS}
In the nearest neighbor search (NNS) problem, data points are represented as $d$-dimensional vectors in $\mathbb{R}^d$, with Euclidean distance $\|p - q\|_2$ as the typical metric. Given a dataset $X = \{x_1, x_2, \dots, x_n\}$, the $k$-nearest neighbor ($k$NN) search for a query $q$ finds the $k$ vectors in $X$ with the smallest distances to $q$. However, in high-dimensional spaces, exact NNS becomes computationally intractable due to the curse of dimensionality~\cite{lsh98}, making approximate nearest neighbor search (ANNS) the practical
standard. An ANNS algorithm returns a result set $R$ with $|R| = k$, evaluated by recall@$k$:
\[
\text{Recall@}k = \frac{|R \cap k\text{NN}(q)|}{k}.
\]

\begin{figure}[!t]
    \centering
    \includegraphics[width=0.80\linewidth]{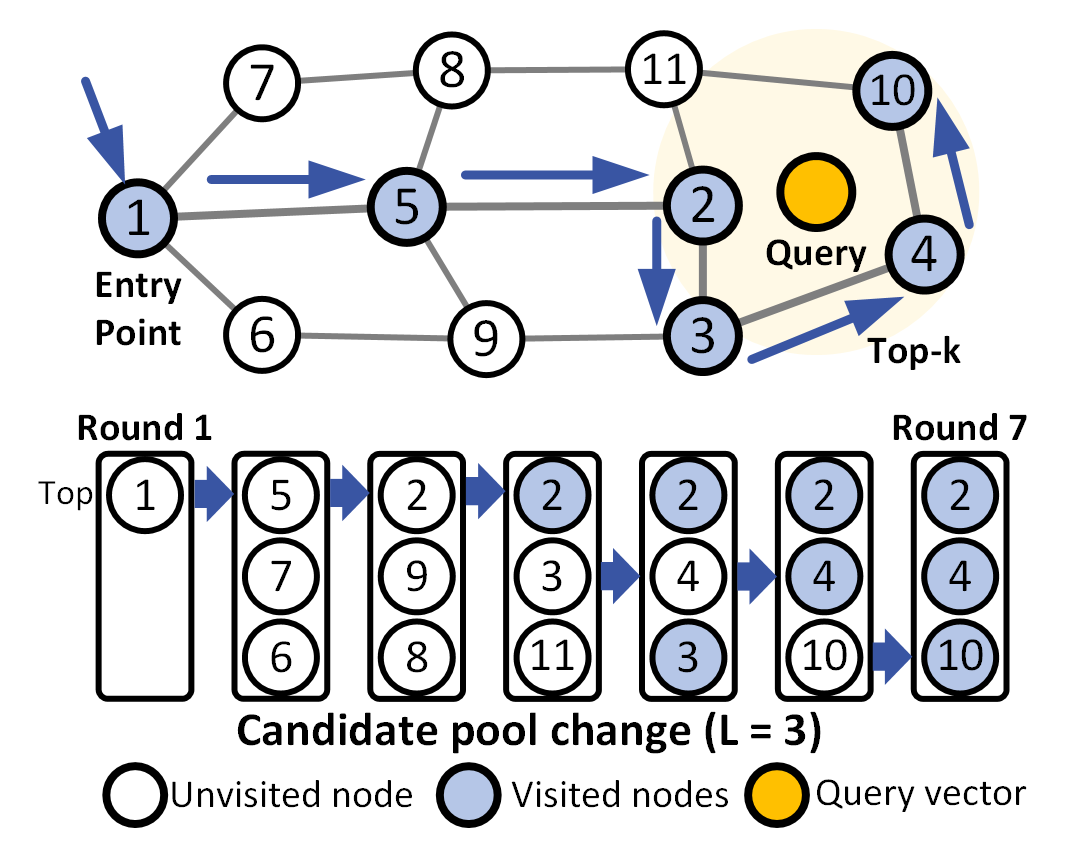}
    \caption{Illustration of greedy search on a proximity graph with candidate pool size $L=3$. The search traverses from the entry point toward the query, updating the candidate pool at each step until convergence.}
    \label{fig:greedy_search}
\end{figure}

Among various ANNS approaches, graph-based methods achieve the best tradeoff between search speed and recall~\cite{diskann, starling}. A graph-based index represents the dataset as a proximity graph $G = (V, E)$, where each vector is connected to its approximate nearest neighbors. As illustrated in Figure~\ref{fig:greedy_search}, given a query $q$, search begins from an entry point (node 1) and maintains a fixed-size min-priority queue $C$ of size $L$ (i.e., \emph{candidate pool}), sorted by distance to the query. At each step, the closest unvisited vector in $C$ is expanded: the distances between its neighbors and the query are computed, and the neighbors are inserted into $C$ ranked by their distances, with vectors beyond the top-$L$ evicted. Initially, only the entry point (node 1) is in the candidate pool, so it is expanded first: its neighbors 5, 7, and 6 are inserted into $C$ ranked by their distances to the query, and node 1 is evicted as it falls beyond the top-$L$. In the next step, the closest unvisited vector (node 5) is expanded, inserting its neighbors 2, 9, and 8 into $C$ and evicting less relevant ones. This process repeats, with the pool progressively converging toward the query, until no unvisited vectors remain in $C$, at which point the top-$k$ vectors in the final candidate pool are returned as the approximate nearest neighbors.

In-memory graph indexes such as HNSW~\cite{hnsw16} and Vamana~\cite{diskann} achieve high recall with sub-millisecond latency. However, in-memory graph-based ANNS requires the entire dataset and graph index to reside in DRAM simultaneously, which incurs substantial memory overhead and limits scalability. To address this, a range of disk-based graph ANNS solutions have been proposed.

\subsection{Disk-based Graph ANNS}
Disk-based ANNS methods offload vectors and graph connections to SSD and read them from disk on demand, while retaining compressed vector representations in memory. Compared to in-memory graph search, disk-based search differs in four key aspects. First, instead of expanding only the single closest unvisited vector per round, disk-based search selects up to $W$ (beam width) closest unvisited vectors per round for expansion. Second, for each selected vector, an I/O request is issued in parallel to fetch its original values and neighbor IDs from SSD. Third, rather than computing exact distances, the approximate distances between the neighbors and the query are computed using their compressed representations already in memory, avoiding additional I/O for getting neighbors' values. Fourth, once the search converges, all vectors visited along the search path have their full-precision values already in memory from prior I/O reads, so a final reranking by full-precision distance is performed and the top-$k$ are returned.

However, this introduces a central challenge: I/O latency. Each traversal step requires reading $W$ pages from SSD, and the random-access nature of graph search makes it difficult to batch or prefetch these reads. Existing methods attack this challenge along four complementary axes.

\noindent\textbf{Disk layout optimization.}
Starling~\cite{starling} and MARGO~\cite{margo} optimize the layout by maximizing mutual neighbor relationships within each page. PageANN~\cite{pageann} aligns graph node granularity with the SSD page size by constructing a page-node graph, eliminating I/O read amplification.

\noindent\textbf{I/O scheduling.}
PipeANN~\cite{pipeann} redesigns the search algorithm to pipeline multiple asynchronous I/O requests, overlapping their latencies to reduce the effective I/O cost per search step. However, compared to standard beam search, pipelining issues suboptimal I/O reads and increases the total I/O count.

\noindent\textbf{Caching frequently visited nodes.}
The reduced memory footprint of disk-based ANNS leaves spare capacity for caching. DiskANN, Starling, MARGO, and PageANN cache the most frequently visited nodes (identified from sample data from datasets) in memory to reduce I/O operations.

\noindent\textbf{In-memory entry point optimization.}
Starling, MARGO, and PipeANN maintain an in-memory graph built over a subset of vectors to reduce the number of disk reads needed to reach the region where nearest neighbors reside.

\section{Motivation}\label{sec:motivation}

Despite the advances mentioned above, existing methods still suffer from severe I/O latency. As shown in Figure~\ref{fig:io_cpu_breakdown}(a), under a memory budget of $0.4\times$ dataset size, where DiskANN caches frequently visited nodes, PipeANN maintains an in-memory graph for entry point optimization, and Starling employs both, I/O latency still accounts for over 90\% of total query latency across all methods and thread counts. Meanwhile, as shown in Figure~\ref{fig:io_cpu_breakdown}(b), the time spent on meaningful CPU work accounts for only 4.7\% to 25.7\% of total query latency across existing methods. Together, these results reveal a severe I/O-CPU imbalance despite existing optimizations: I/O dominates query latency while the CPU remains largely idle for the vast majority of query time.

\begin{figure}[!t]
    \centering
    \includegraphics[width=0.90\linewidth]{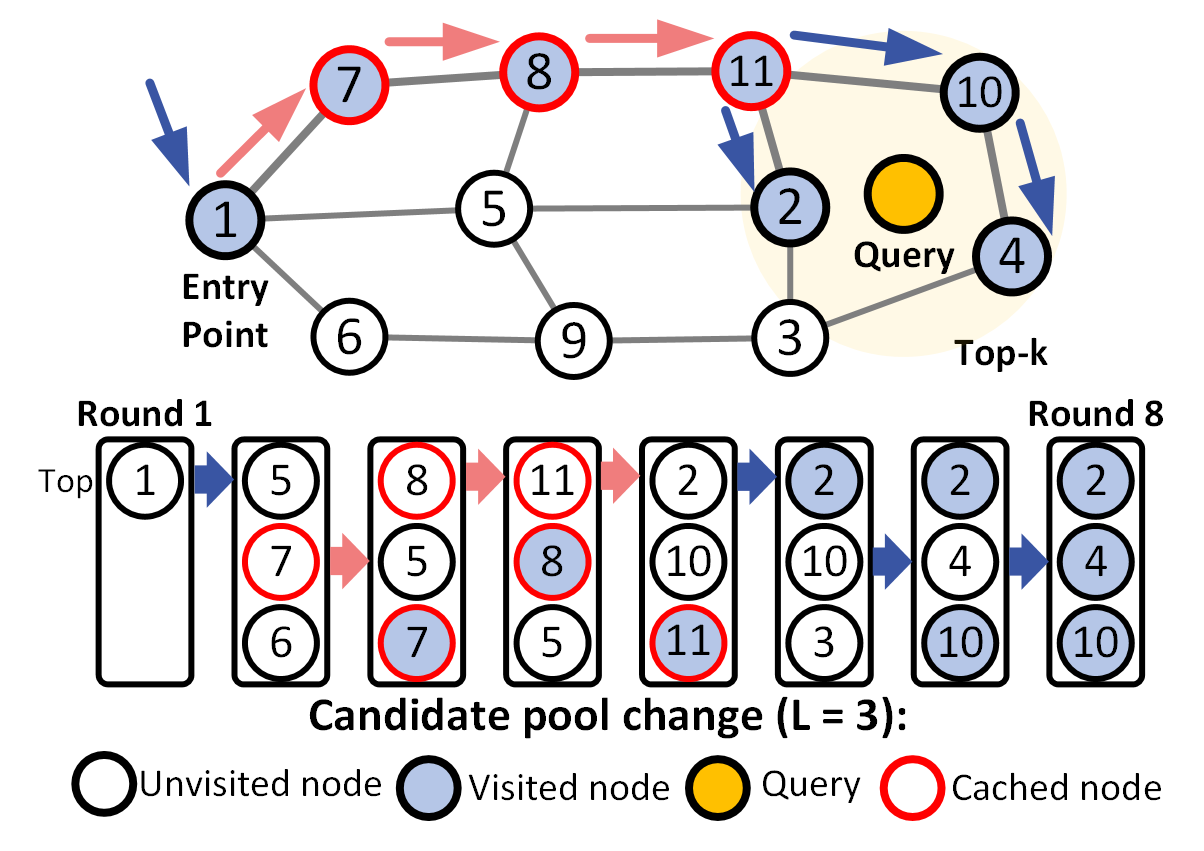}
   \caption{Illustration of processing vectors cached in memory before issuing I/O on a proximity graph with candidate pool size $L=3$. Blue arrows indicate expansions that require I/O; red arrows indicate expansions of vectors cached in memory requiring no I/O. Red-circled nodes are cached in memory.
}
\label{fig:mem_first_search}
\end{figure}

However, CPU work and the number of I/O operations are tightly coupled. As discussed in Section~\ref{sec:background}, existing methods already store frequently visited nodes in memory using spare memory capacity. Processing these in-memory vectors before making I/O decisions in each round can potentially reduce the total number of I/O operations for the following two reasons.

First, because graph search often contains multiple paths from the entry point to the nearest-neighbor region, expanding a non-closest unvisited vector that is already in memory can still advance the search frontier and reduce the number of disk reads needed to reach the top-$k$ region. Figure~\ref{fig:mem_first_search} illustrates this effect. In the second round, instead of issuing an I/O for node 5, the closest unvisited vector, the search expands node 7, which is already in memory. This expansion inserts node 8 into the candidate pool, and node 8 is closer to the query than node 5. As a result, node 5 is bypassed and eventually evicted before its I/O is issued. The same process continues through nodes 8 and 11, allowing the search frontier to move toward the query without additional disk reads. In this way, newly discovered closer candidates progressively displace less relevant ones in the candidate pool, causing nodes such as 5 and 3 to be evicted before they trigger I/O and thereby avoiding disk accesses that would not contribute to the final result.

Moreover, disk-based graph traversal ranks candidates using approximate distances from compressed vectors, rather than exact distances. This makes the ranking imperfect: a candidate that appears lower in the pool may actually be closer to the query when its exact distance is computed. Standard greedy beam search may evict such candidates before expanding them, missing useful paths toward the true nearest neighbors. By processing these lower-ranked but already available candidates, the search can recover such missed opportunities, improve final accuracy, and reduce the number of I/Os needed to reach a given recall target.

\begin{figure}[!t]
    \centering
    \includegraphics[width=0.99\linewidth]{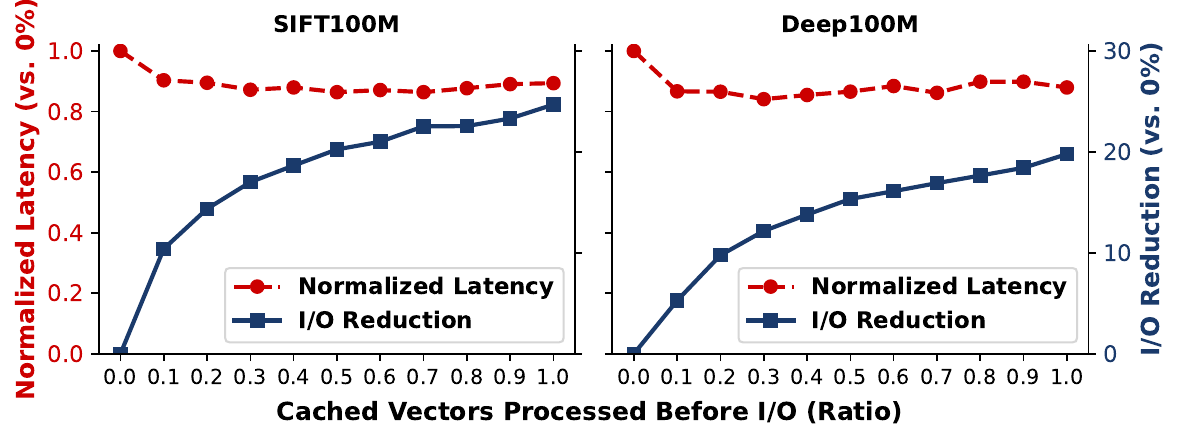}
    \caption{Normalized latency and I/O reduction (relative to 0 ratio) as a function of the ratio of in-memory vectors processed before issuing I/O per round, on SIFT100M and Deep100M with DiskANN.}
    \label{fig:cpu_work_reduces_io}
\end{figure}

As shown in Figure~\ref{fig:cpu_work_reduces_io}, before issuing I/O for the next round, we identify all unexplored candidates in the search pool whose vectors are available in memory and expand the top $X\%$ ranked by approximate distance to the query. As $X$ increases, the total number of I/O operations decreases consistently on both datasets. This result confirms our observation: additional CPU processing on already available candidates can guide the search more effectively and reduce subsequent disk accesses.

\textbf{Challenge: balancing CPU work and I/O progress.} 
Although processing more available candidates can reduce disk accesses, it also introduces a fundamental tradeoff. Additional candidate processing increases CPU overhead in each search round and may delay I/O submission for the top-$L$ candidates that remain in the search pool. As shown in Figure~\ref{fig:cpu_work_reduces_io}, query latency initially decreases as more available candidates are processed, but eventually plateaus or increases beyond a certain processing ratio. This trend is caused by two factors. First, the marginal benefit of I/O reduction diminishes as increasingly less relevant candidates are processed, while CPU overhead continues to grow. Second, delaying I/O submission for high-priority candidates directly prolongs query latency, because their full-precision vectors must be fetched before the search can complete. Table~\ref{tab:beamwidth} further confirms this effect: although a smaller beam width $W$ reduces the total number of I/Os, it postpones I/O issuance and leads to significantly higher latency than a larger $W$. The key challenge is therefore to reduce I/Os through additional candidate processing while limiting CPU overhead and preserving timely I/O submission for high-priority candidates.

A straightforward approach is to move this additional CPU work into I/O waiting periods, when the CPU would otherwise remain largely idle, as shown in Figure~\ref{fig:io_cpu_breakdown}(b). However, exploiting this idle time effectively is challenging for two reasons. First, not all available candidates can be safely deferred. Some candidates, once expanded, may reveal neighbors that become the closest unvisited candidates in the search pool and therefore directly affect the next I/O decision. If these candidates are processed only after I/O requests have been issued, the system makes I/O decisions based on a stale pool state, which can trigger unnecessary disk reads and reduce I/O efficiency. Second, the amount of useful CPU work available during each I/O wait is limited. Because ANNS maintains a relatively small candidate pool to keep query latency low, many available candidates are quickly evicted before they can be processed. As a result, the CPU may still become idle during much of the I/O wait, limiting both the opportunity to reduce I/Os and the potential accuracy gains from additional candidate processing.

\begin{table}[!t]
\centering
\small
\caption{Impact of beam width $W$ on total \#I/Os, latency,
and QPS on SIFT100M with DiskANN at Recall@10 = 0.9.}
\label{tab:beamwidth}
\resizebox{\columnwidth}{!}{%
\begin{tabular}{lccccc}
\toprule
Beam width $W$ & 1 & 2 & 4 & 8 & 16 \\
\midrule
Total \#I/Os  & \textbf{183.73} & 187.99 & 196.26 & 214.87 & 252.37 \\
Latency (ms)  & 24.50 & 15.24 & \textbf{11.26} & 12.15 & 13.78 \\
QPS           & 325.79 & 523.42 & \textbf{707.35} & 655.51 & 577.79 \\
\bottomrule
\end{tabular}%
}
\end{table}


\section{LAANN Design}\label{sec:design}
In this section, we introduce our proposed LAANN, that makes graph search explicitly I/O-aware by co-optimizing CPU computation and I/O access.
\subsection{System Overview}

\begin{figure*}[!t]
    \centering
    \includegraphics[width=0.999\textwidth]{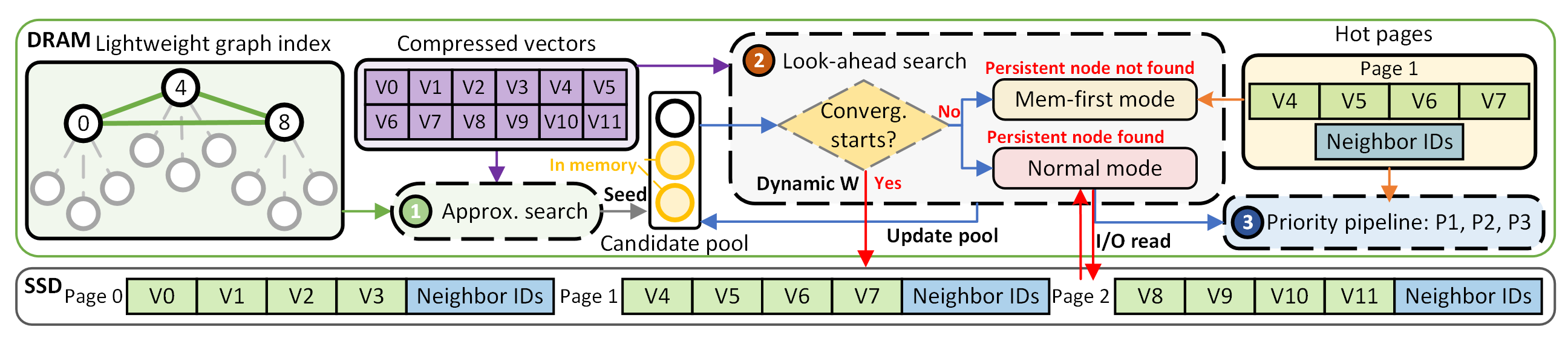}
    \caption{System overview of LAANN. The DRAM region holds three purpose-built structures; the SSD stores the page-level disk graph. Numbers indicate the search flow.}
    \label{fig:system_overview}
\end{figure*}

LAANN addresses the I/O bottleneck through three complementary techniques. First, a \emph{look-ahead search} adapts its strategy to the current search phase to balance I/O reduction and timely I/O issuing. Second, a \emph{priority I/O-CPU pipeline} exploits CPU idle time during I/O waiting to reduce CPU overhead, I/Os, and improve accuracy. Third, a \emph{fast lightweight in-memory graph index} further reduces I/Os by seeding the disk graph search with high-quality candidates from the start. To support these, LAANN partitions the available memory budget into three structures, as illustrated in Figure~\ref{fig:system_overview}: \emph{compressed vectors} for approximate distance computation, an \emph{lightweight in-memory graph index} representing edge connections among centroid vectors of each page node, and \emph{cached disk pages} containing frequently accessed disk graph nodes each corresponding to an SSD page.

The search proceeds in three steps. First, the lightweight in-memory graph index (\circled{1}) traverses entirely in memory and seeds the disk graph candidate pool with high-quality candidates (Section~\ref{sec:highway}). Second, the look-ahead search (\circled{2}) takes over and adapts its strategy to the current phase: in the approach phase, it alternates between memory-first mode and normal mode via a persistence check; in the convergence phase, it applies a dynamic beam width scheme to minimize delays in issuing I/Os for vectors that remain in the final candidate pool (Section~\ref{sec:lookahead}). Third, the priority I/O-CPU pipeline (\circled{3}) fills CPU idle time with computation ordered by relevance to the next I/O decision, supported by a customized candidate pool with an overflow area (Section~\ref{sec:pipeline}).

\subsection{Look-Ahead Search}
\label{sec:lookahead}

As discussed in Section~\ref{sec:motivation}, naively processing more in-memory vectors introduces two conflicting costs that offset the I/O reduction benefit: additional CPU overhead per round, and delayed I/O submission for the top-$L$ vectors that remain in the final candidate pool, which directly prolongs query latency. To address this, we need to dive deep into the I/O composition of disk-based graph search.

I/Os in disk-based graph search can be generally classified into two types: I/Os for vectors that remain in the final candidate pool, whose count is largely predetermined by the target recall accuracy; and I/Os for vectors that do not appear in the final candidate pool upon convergence, issued to navigate the graph toward the final pool vectors, whose count can vary largely depending on the search path taken. At the same time, prior studies~\cite{pipeann, govector} have shown that disk-based graph search can be divided into two distinct phases. In the \emph{approach phase}, the search traverses from the entry point toward the region where the nearest neighbors reside. In the \emph{convergence phase}, the search refines locally around the nearest neighbors until the whole candidate pool stabilizes.

\begin{figure}[!t]
    \centering
    \includegraphics[width=0.99\linewidth]{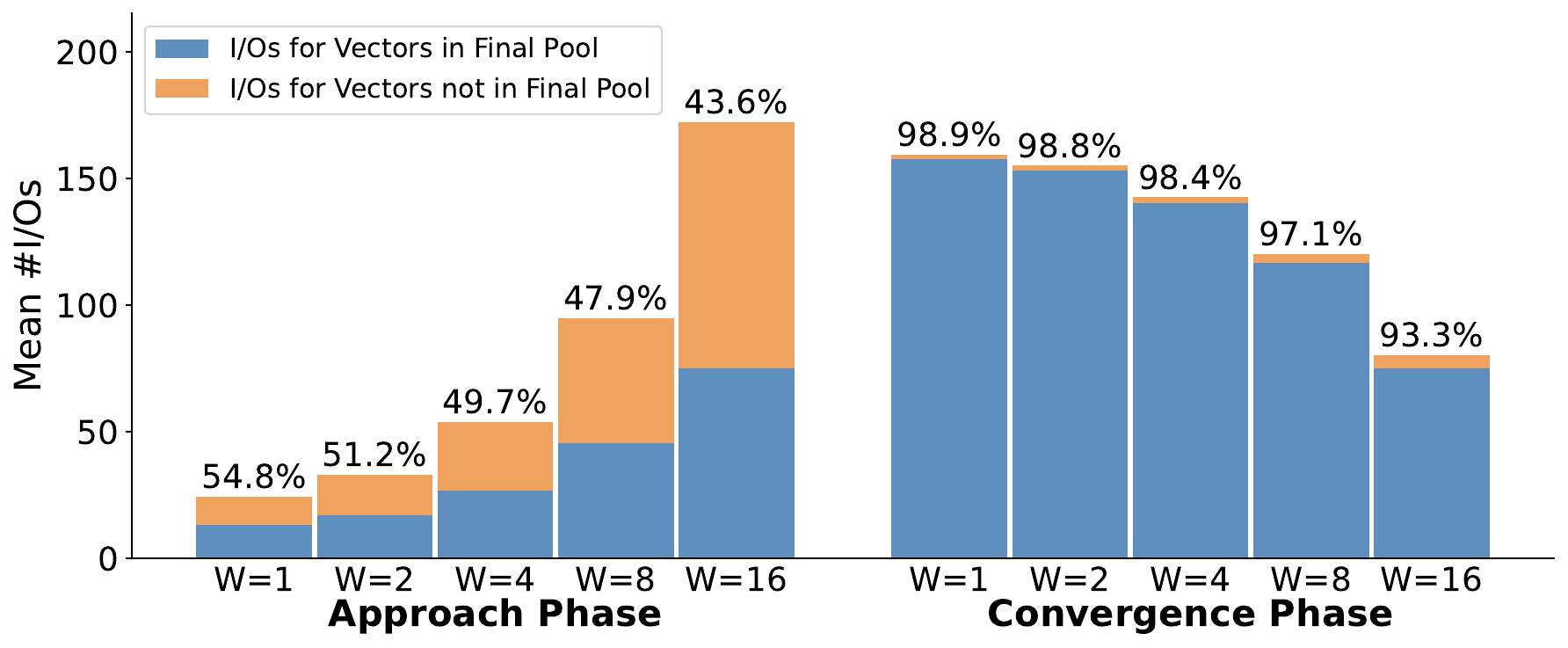}
    \caption{I/O composition across the approach phase and convergence phase under different beam widths $W$, on SIFT100M with DiskANN. Each bar shows the mean \#I/Os broken down into I/Os for vectors that remain in the final candidate pool and I/Os for vectors that do not. Percentages indicate the fraction of I/Os for vectors in the final pool.}
    \label{fig:io_waste}
\end{figure}

As illustrated in Figure~\ref{fig:io_waste}, the two phases exhibit fundamentally different I/O compositions. In the approach phase, the candidate pool is rapidly changing and most visited vectors are intermediate nodes that do not survive to the final pool. Only 43.6\%--54.8\% of the I/Os issued during this phase are for vectors in the final pool, meaning the majority can potentially be reduced. In the convergence phase, the top-$L$ approximate nearest neighbors have largely been identified and the pool becomes stable. 93.3\%--98.9\% of the I/Os issued during this phase are for vectors in the final pool, meaning the vast majority are essential for returning the final results.

These observations suggest that the optimal search strategy differs fundamentally between the two phases. In the approach phase, where the majority of I/Os are for vectors not in the final pool, the search should be more conservative about issuing I/Os and prioritize processing in-memory neighbors instead. In the convergence phase, where the vast majority of I/Os are for vectors in the final pool, the search should instead issue I/Os more aggressively to minimize the delay in receiving their full-precision values. Based on these observations, we propose \emph{look-ahead search}, a novel search algorithm that adapts its search strategy to the current phase.

\begin{figure}[!t]
    \centering
    \includegraphics[width=0.95\linewidth]{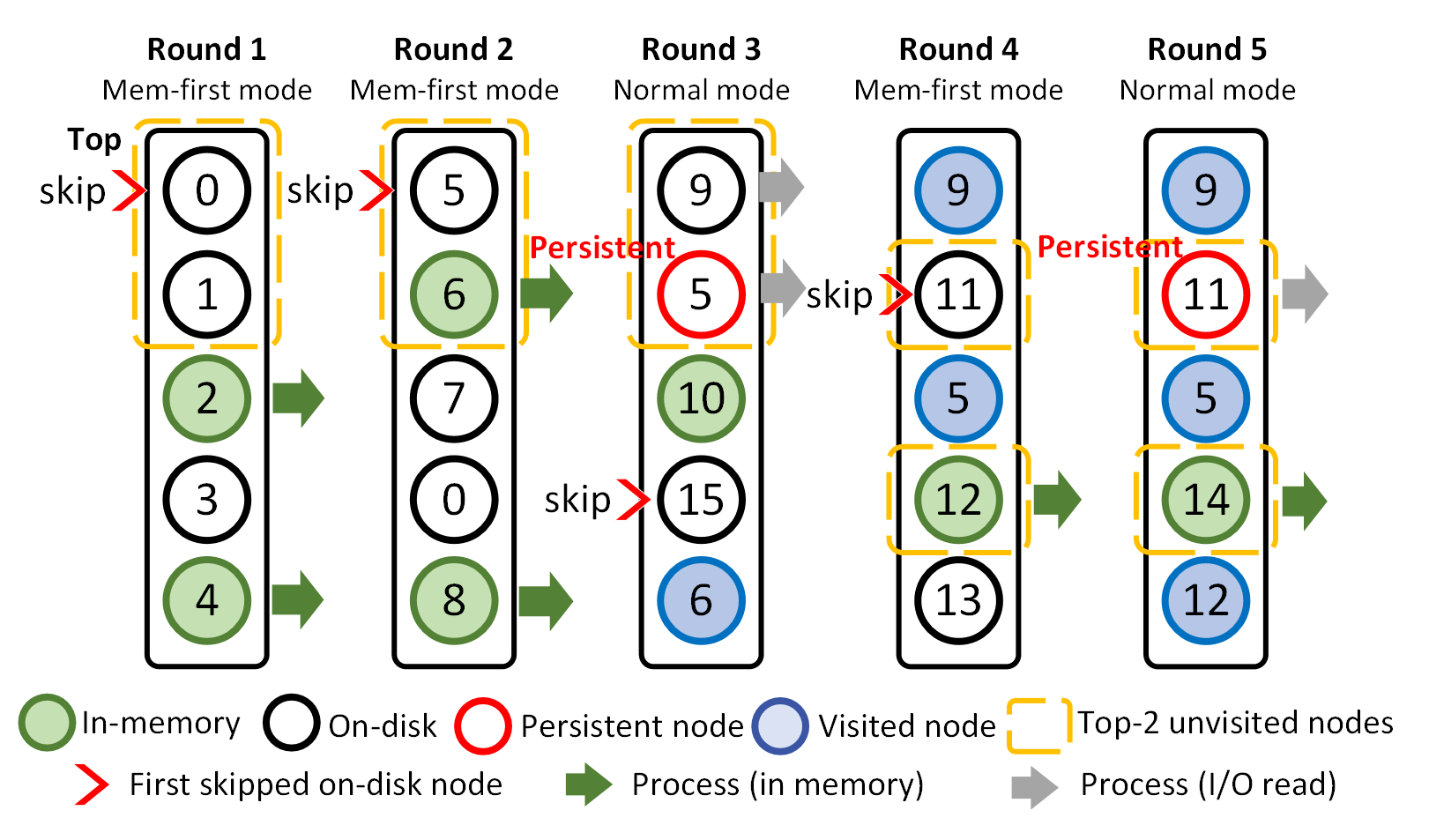}
    \caption{Look-ahead search with $W = 2$. Green nodes are cached in memory; white nodes are on-disk; blue nodes have been explored. Yellow dashed boxes mark the top-$W$ unvisited vectors inspected by the persistence check. Red circles indicate persistent nodes.}
    \label{fig:look_ahead}
\end{figure}

\noindent\textbf{Look-Ahead Search in the Approach Phase.} Following Pipe-ANN~\cite{pipeann}, we detect the convergence phase by checking whether all top-$n$ vectors in the candidate pool have been explored, where $n \ll L$. Once this condition is met, the search transitions from the approach phase to the convergence phase. During the approach phase, the look-ahead search alternates between two modes: \emph{memory-first mode} and \emph{normal mode}, as illustrated in Figure~\ref{fig:look_ahead} and detailed in Algorithm~\ref{alg:lookahead_search}.

In memory-first mode, the algorithm scans the candidate pool in ascending distance order and collects up to $W$ in-memory vectors to explore, skipping over on-disk vectors. The first skipped on-disk vector is recorded as $\textit{skipped}$ for the next round. By prioritizing in-memory vectors, this mode reduces I/Os issued for on-disk vectors that are not in the final candidate pool.

In normal mode, the algorithm falls back to standard greedy beam search behavior: it collects up to $W$ vectors from the candidate pool regardless of residency, issuing I/Os for on-disk vectors as needed. The next closest unvisited on-disk vector remaining in the candidate pool is recorded as $\textit{skipped}$ for the next round.

The switching between the two modes is governed by a \emph{persistence check} performed at the beginning of each round. The algorithm inspects the top $W$ unvisited vectors in the candidate pool to check whether $\textit{skipped}$, the first on-disk vector skipped in the previous round, still appears within this window. If so, the skipped vector is considered \emph{persistent}: no closer in-memory neighbor was inserted to push it out of the window, indicating it is critical to the search progress and must be fetched from disk. In this case, the search switches to normal mode for the current round. Otherwise, the search continues in memory-first mode. As shown in Figure~\ref{fig:look_ahead}, in round~1, on-disk node~0 is skipped and recorded as $\textit{skipped}$, while in-memory nodes~2 and~4 are processed. In round~2, node~0 no longer appears within the top-$W$ unvisited nodes, so memory-first mode continues: node~5 becomes the new $\textit{skipped}$, and in-memory nodes~6 and~8 are processed. In round~3, node~5 remains persistent within the top-$W$ unvisited vectors, so the search switches to normal mode and expands all top-$W$ unvisited vectors regardless of residency, recording the next closest unvisited on-disk vector as $\textit{skipped}$ for the following round.

\begin{figure}[!t]
    \centering
    \includegraphics[width=0.99\linewidth]{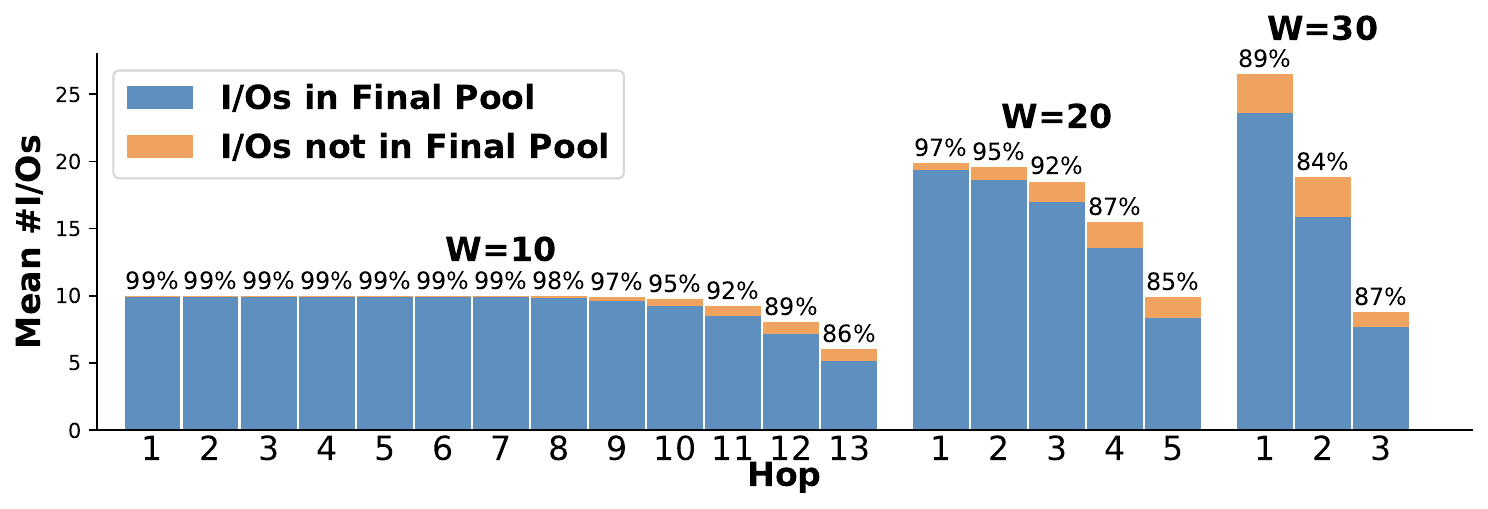}
    \caption{I/O composition at each hop during the convergence phase under different beam widths $W$, on SIFT100M with DiskANN. Each bar shows the mean \#I/Os broken down into I/Os for vectors in the final pool and I/Os for vectors not in the final pool. Percentages indicate the fraction of I/Os for vectors in the final pool.}
    \label{fig:convergence_io_composition}
\end{figure}

\noindent\textbf{Look-Ahead Search in the Convergence
Phase.} Once the convergence phase is detected, the search shifts its priority to minimizing the delay in issuing I/Os for vectors in the final candidate pool. As shown in Figure~\ref{fig:convergence_io_composition}, although the majority of I/Os in the convergence phase are for vectors in the final pool, vectors near the end of the candidate pool still have a chance of being evicted as new neighbors are inserted, and this eviction probability increases as exploration progresses closer to the end of the pool. The key is therefore to issue I/Os aggressively for vectors near the top of the pool, while being progressively more conservative toward the end.

PipeANN, however, increases the beam width by one at each round once convergence is detected. This rigid linear increase does not account for the increasing probability of vectors being evicted from the final candidate pool as exploration progresses closer to the end of the pool, resulting in overly conservative I/O issuing at the start of the convergence phase and overly aggressive issuing near the end, causing both I/O waste and I/O issuing delays.

Instead, we propose a dynamic beam width scheme that spikes aggressively at the start of the convergence phase and then decays gradually. Specifically, the beam width $W_{conv}$ is initialized to $\alpha \times L$ in the first round, where $\alpha \in (0, 1)$ is the initial spike ratio relative to the candidate pool size $L$, providing an aggressive initial burst of I/O issuing. In each subsequent round, $W_{conv}$ is updated as:
\begin{equation}
W_{conv} \leftarrow \max\left(\lfloor W_{conv} \times
\beta \rfloor,\ W\right)
\end{equation}
where $\beta \in (0, 1)$ is the decay ratio that progressively reduces $W_{conv}$ each round, making the search more conservative about issuing I/Os for vectors near the end of the pool as exploration progresses. This continues until all vectors within the top-$L$ of the candidate pool have been visited. A larger $\alpha$ issues I/Os more aggressively at the start but risks I/O waste near the end of the pool; $\beta$ mitigates this by decaying $W_{conv}$ progressively, becoming more conservative as exploration approaches the end of the pool. Together, they minimize both I/O issuing delay for vectors in the final candidate pool and I/O waste. Based on our experiments, we recommend $\alpha = 0.25$ and $\beta = 0.95$.

\begin{algorithm}[!t]
\caption{LAANN Look-Ahead Search}
\label{alg:lookahead_search}
\begin{algorithmic}[1]
\Require Candidate pool $C$ with $|C| = \mu \times L$,
         disk graph $G$, query $q$,
         beam width $W$, stability threshold $n$,
         initial spike ratio $\alpha$, decay ratio $\beta$
\Ensure Top-$k$ nearest neighbors of $q$
\State $\textit{skipped} \gets \bot$;\
       $\textit{converged} \gets \textsc{False}$;\
       $W_{conv} \gets \bot$
\While{$C$ has unvisited nodes}
    \If{\textbf{not} $\textit{converged}$}
        \State $\textit{converged} \gets \textsc{CheckConvergence}(C, n)$
    \EndIf
    \If{$\textit{converged}$}
        \State $W_{conv} \gets \textsc{UpdateBeamWidth}(W_{conv}, \alpha, \beta, L, W)$
        \State Collect all unvisited vectors within top-$W_{conv}$ of $C$
    \ElsIf{$\textit{skipped} \neq \bot$ \textbf{and}
           $\textit{skipped} \in$ top-$W$ unvisited in $C$}
        \State Collect top-$W$ unvisited vectors from $C$
               regardless of residency
        \State $\textit{skipped} \gets$ next unvisited on-disk
               vector in $C$, or $\bot$
    \Else
        \State Collect top-$W$ in-memory vectors from $C$
        \State $\textit{skipped} \gets$ top unvisited on-disk
                vector in $C$, or $\bot$
    \EndIf
    \State Issue async I/O for collected on-disk vectors;
           mark all collected as visited
    \State Insert neighbors of collected vectors into $C$
\EndWhile
\State \Return Top-$k$ from $C$ ranked by full-precision distance
\Statex
\Function{CheckConvergence}{$C$, $n$}
    \State \Return $n$-th vector in $C$ is unchanged from
           last round
\EndFunction
\Statex
\Function{UpdateBeamWidth}{$W_{conv}$, $\alpha$, $\beta$, $L$, $W$}
    \If{$W_{conv} = \bot$} \Return $\lfloor \alpha \times L \rfloor$
    \Else\ \Return $\max\left(\lfloor W_{conv} \times \beta \rfloor,\ W\right)$
    \EndIf
\EndFunction
\end{algorithmic}
\end{algorithm}

\subsection{Priority I/O-CPU Pipeline with Customized Candidate Pool}
\label{sec:pipeline}

\begin{figure}[!t]
    \centering
    \includegraphics[width=0.99\linewidth]{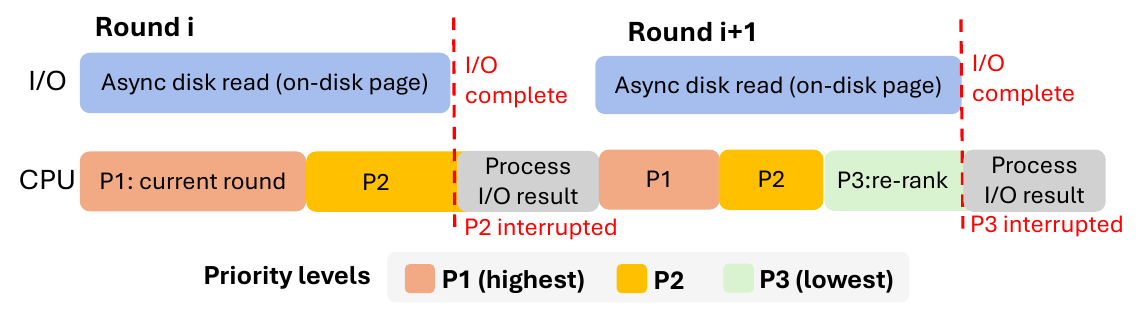}
    \caption{Priority I/O-CPU pipeline across two consecutive
    search rounds. Lower-priority tasks are interrupted upon
    I/O completion. P1: compute approximate distances for
    neighbors of in-memory vectors selected in the current
    round; P2: expand in-memory vectors outside the current
    search window to improve search accuracy; P3: re-rank
    visited vectors by full-precision distances for final
    top-$k$ results.}
    \label{fig:pipeline}
\end{figure}

As discussed in Section~\ref{sec:motivation}, the CPU remains idle for the vast majority of I/O waiting time in existing methods. Moreover, as discussed above, processing in-memory vectors beyond the closest ones can improve search accuracy and thus reduce the number of I/Os needed to achieve a given recall target. We therefore propose a \emph{priority I/O-CPU pipeline} that systematically fills CPU idle time during I/O waiting with useful work, ordered by relevance to the next I/O decision. Figure~\ref{fig:pipeline} illustrates
the pipeline across two consecutive rounds.

\noindent\textbf{Priority levels.} All deferrable CPU tasks are classified into three priority levels based on their relevance to the next I/O decision.

\emph{Priority 1 (highest):} Computing approximate distances for the neighbors of the in-memory vectors selected for expansion in the current round. Since these vectors are selected for expansion, the distance computation and insertion of their neighbors directly determine the next state of the candidate pool and the next I/O decision. Hence, as shown in Figure~\ref{fig:pipeline}, P1 always executes first and is non-interruptible, as delaying it would cause I/O decisions to be made on a stale pool state.

\emph{Priority 2:} Expanding in-memory vectors present in the candidate pool $C$ but not selected for expansion in the current round. Unlike P1, their neighbors are less directly relevant to the next I/O decision. However, processing them can potentially update the candidate pool, reduce future I/Os, and improve final search accuracy. LAANN processes these vectors one at a time during I/O waiting time, checking after each whether the pending I/O has completed and halting immediately upon completion to avoid delaying the search.

\emph{Priority 3 (lowest):} Computing full-precision distances for all vectors visited throughout the search and ranking them accordingly. These computations have no relevance to I/O decisions and are entirely independent of I/O issuing. Unlike traditional methods which perform reranking after the search terminates, LAANN incrementally computes and maintains the ranking of all visited vectors by full-precision distance during I/O waiting time once P1 and P2 work is exhausted, leaving minimal work after the search terminates.

\begin{figure}[!t]
    \centering
    \includegraphics[width=0.95\linewidth]{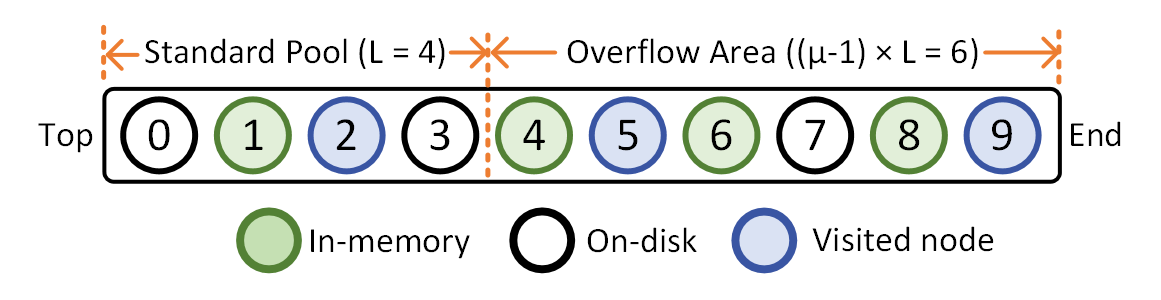}
    \caption{Candidate pool with overflow area showing $\mu$ × L = 10 total positions with $\mu$ = 2.5 and L = 4 (4 standard pool + 6 overflow area). 
    }
    \label{fig:candidate_pool_overflow}
\end{figure}


\noindent\textbf{Candidate pool with overflow area.} As discussed in Section~\ref{sec:motivation}, the supply of unvisited in-memory vectors within the candidate pool is quickly exhausted during I/O waiting time, leaving the CPU idle and limiting the opportunity to reduce I/Os and improve accuracy. To address this, we enlarge the candidate pool to $\mu \times L$ while keeping the convergence condition unchanged: the search still terminates when the top-$L$ vectors are fully visited. As illustrated in Figure~\ref{fig:candidate_pool_overflow}, the extra $(\mu - 1) \times L$ positions act as an overflow area that serves as a natural reservoir of in-memory vectors (shown in green) for P2 work during I/O waiting time. Since vectors are inserted into the overflow area through the normal pool insertion process, they are already ranked by approximate distance to the query, requiring no extra computation to assess their relevance. The figure shows how in-memory vectors are distributed throughout both the standard pool and overflow area, providing P2 with a continuous supply of work opportunities. A larger $\mu$ provides more P2 work supply but contributes diminishing returns beyond a certain point while increasing pool management overhead; $\mu \in [2, 3]$ provides a good balance in practice. The overflow area requires no extra data structure, adds negligible overhead, and does not change the priority logic, making it a lightweight co-design between the search algorithm and the pipeline.

\noindent\textbf{Effect.} By ordering CPU work during I/O waiting time by its relevance to the next I/O decision, the pipeline reduces I/Os, improves search accuracy, and overlaps the majority of deferrable computation with I/O waiting time. The amount of work completed adapts automatically to the variable latency of each I/O request: shorter waits complete only P1, longer waits additionally execute P2 and P3. The overflow area ensures a sustained supply of useful P2 work throughout the search, keeping the CPU productive during I/O waiting time at negligible overhead.

\subsection{Lightweight In-Memory Graph Index}
\label{sec:highway}

\begin{figure}[!t]
    \centering
    \includegraphics[width=0.99\linewidth]{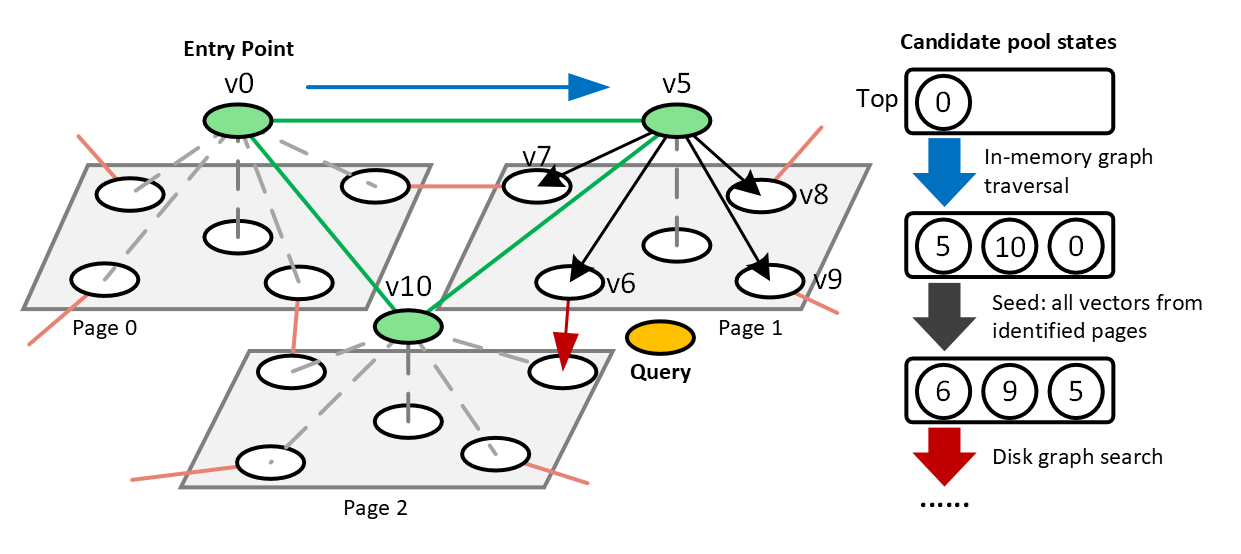}
    \caption{Lightweight in-memory graph index search and disk graph seeding. Green nodes are per-page centroids; green edges are index connections. The right side shows the candidate pool evolution across the three steps: index search, seeding, and disk graph search.}
    \label{fig:highway_network}
\end{figure}

\noindent\textbf{Mismatch in existing in-memory graph indexes.} Existing methods~\cite{starling, margo, pipeann}, inspired by HNSW~\cite{hnsw16}, build in-memory graph indexes by randomly sampling vectors from the dataset and constructing a Vamana graph over them. During search, they traverse these indexes based on full-precision distances, which requires loading the raw vector values of the sampled vectors into memory. However, disk graph traversal is driven by approximate distances from compressed vectors. The entry points found by full-precision search may not be the closest vectors for a search that ranks by approximate distance. Moreover, due to this precision mismatch, these entry points can only serve as starting positions and cannot be directly added to the disk graph candidate pool, forcing the disk graph search to start from scratch with a nearly empty candidate pool.

\noindent\textbf{Key insight.} To approach the regions where the nearest neighbors reside, searching by approximate distance is not only sufficient, but also significantly faster and more memory-efficient than full-precision distance computation. Furthermore, in order to converge faster, what the disk graph search truly needs is a candidate pool filled with neighbors close to the query based on their approximate distances, rather than a few entry points to start the search from a nearly empty pool.

\noindent\textbf{Index Construction.} LAANN builds its lightweight in-memory graph index on top of the page-aligned disk graph design of Page-ANN~\cite{pageann}, which groups spatially close vectors into the same disk page and treats each page as a single graph node. The lightweight in-memory graph index is a Vamana graph built over the centroid vector of each disk page. As shown in Figure~\ref{fig:highway_network}, each page groups spatially close vectors and its centroid naturally represents the local region covered by that page. The index connects these centroids with long-range edges, enabling the search to traverse large graph distances in few hops without accessing the disk. Since the index search uses compressed representations already in memory for distance computation, it requires no additional memory overhead for vector storage. When the memory budget is sufficient, one centroid per disk page is used, providing full graph coverage; when constrained, a random subset of centroids is sampled to reduce the memory footprint. LAANN's look-ahead search and priority I/O-CPU pipeline are built on top of this disk graph design and are readily transferable to other disk-based ANNS systems.

\noindent\textbf{Seeding the Disk Graph Candidate Pool.} As illustrated in Figure~\ref{fig:highway_network}, the lightweight in-memory graph index search begins from an entry centroid and traverses the index entirely in memory, identifying the region where the nearest neighbors reside without issuing any disk I/O. Once the search converges, LAANN seeds the disk graph candidate pool by iterating through all centroids in the result set, computing the page ID of each centroid, deriving the vector IDs within the page based on the page ID, computing their approximate distances to the query, and inserting them into the disk graph candidate pool. This transforms a pool of centroid candidates into a pool of high-quality vector candidates concentrated near the true nearest neighbors. Since both the index search and the disk graph search use the same approximate distance metric, there is no precision mismatch and the seeded candidates can be directly used in the disk graph traversal. The disk graph search then starts directly from this pre-filled candidate pool, reducing the number of I/Os needed to navigate toward the nearest neighbors in the approach phase.

\begin{algorithm}[!t]
\caption{Lightweight In-Memory Graph Index Search and Seeding}
\label{alg:highway_search}
\begin{algorithmic}[1]
\Require Lightweight in-memory graph index $G_a$ with degree $R_a$,
         query $q$, index pool size $L_a$,
         disk graph candidate pool $C$
\Ensure $C$ initialized with high-quality candidates
\State $C_a \gets$ empty priority queue of capacity $L_a$
\State $s \gets$ medoid of $G_a$
\State Insert $(s,\ \tilde{d}(s, q))$ into $C_a$
\While{$C_a$ has unvisited nodes}
    \State $v \gets$ closest unvisited node in $C_a$ by $\tilde{d}(v, q)$
    \ForAll{neighbor $u \in G_a.\textit{neighbors}(v)$}
        \State Insert $(u,\ \tilde{d}(u, q))$ into $C_a$
    \EndFor
\EndWhile
\State \Comment{Seed disk graph candidate pool}
\State $V_p \gets \emptyset$ \Comment{Visited page IDs}
\ForAll{node $v \in C_a$}
    \State $p \gets \textit{pageID}(v)$
    \If{$p \notin V_p$}
        \State $V_p \gets V_p \cup \{p\}$
        \ForAll{vector $u$ in page $p$}
            \State Insert $(u,\ \tilde{d}(u, q))$ into $C$
        \EndFor
    \EndIf
\EndFor
\end{algorithmic}
\end{algorithm}


\section{Implementation}\label{sec:implementation}

LAANN is implemented in C++ with roughly 4K lines of code on top of the PageANN codebase. LAANN adopts the disk graph design of PageANN~\cite{pageann}, which groups the closest vectors to a centroid into the same disk page and treats each page as a single graph node. This page-aligned design matches each graph node to exactly one SSD page, eliminating I/O read amplification and enabling the seeding process to efficiently expand centroid candidates to all vectors within a page.

\noindent\textbf{Asynchronous I/O.}
LAANN uses Linux \texttt{io\_uring} for asynchronous disk reads. When the look-ahead search enters normal mode, I/O requests for on-disk pages are submitted to the \texttt{io\_uring} submission queue in a single batch via \texttt{io\_uring\_submit}. While the priority pipeline executes CPU work during the I/O wait, completed requests are collected via non-blocking \texttt{io\_uring\_peek\_batch\_cqe} polls between priority tasks. If no deferrable CPU work remains, the search blocks on \texttt{io\_uring\_wait\_cqe} until all I/O complete. This design ensures that the CPU processes priority tasks during I/O latency and blocks only when no useful work is available.

\noindent\textbf{Cached page loading.} LAANN profiles page visit frequencies by running the lightweight in-memory graph index search on a 1\% sample of the dataset, ranks page nodes by their visit frequency, and writes the ordering to a file. At load time, page nodes are loaded into memory following this ordering, ensuring the most frequently visited pages are loaded first. Residency is checked via a hash table at query time. Optionally, pages can be physically reordered on disk according to this file, eliminating the memory overhead of the hash table: residency can then be determined by a simple index comparison, since a page is cached if and only if its index is below the number of loaded pages.


\begin{figure*}[!t]
    \centering
    \includegraphics[width=0.99\linewidth]{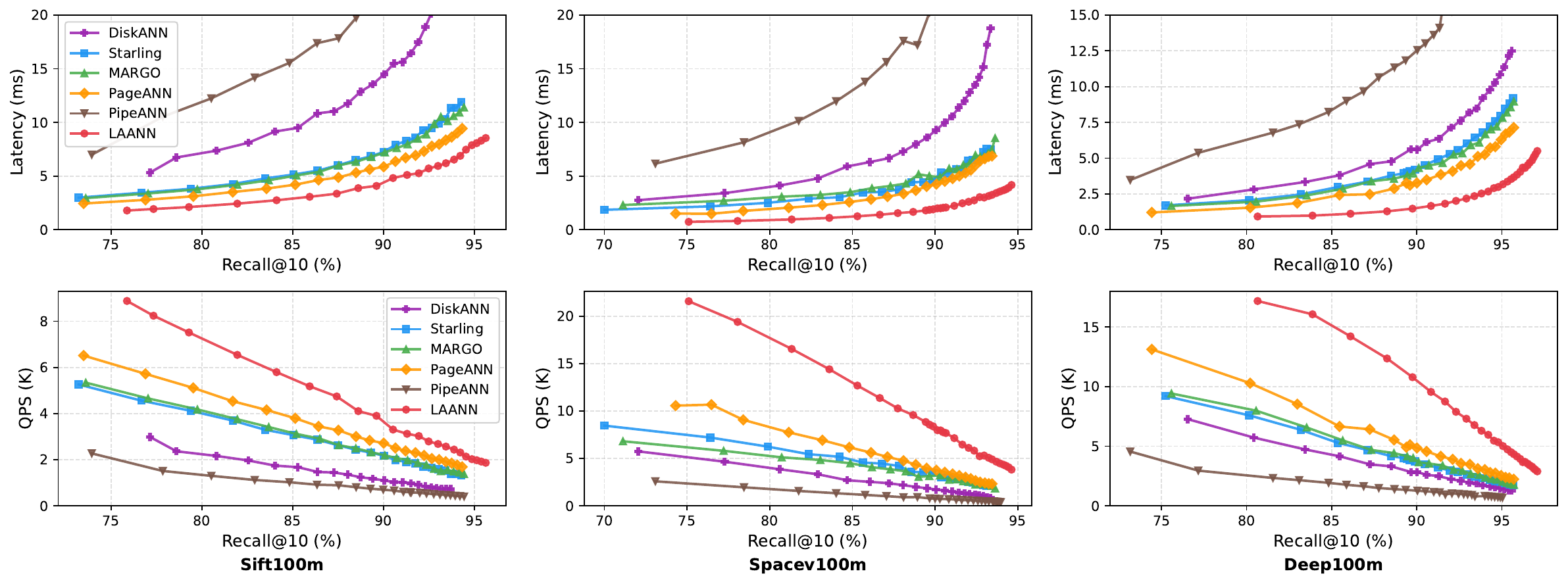}
    \caption{Recall@10 vs. latency (top row) and recall@10 vs. throughput (bottom row) on three 100M-scale datasets under a memory budget of $0.5\times$ the dataset size. Lower latency and higher throughput are better.}
    \label{fig:overall_performance_100m}
\end{figure*}

\section{Evaluation}\label{sec:results}

In this section, we evaluate the performance of LAANN against SOTA disk-based ANNS schemes under various workloads and memory budgets.

\subsection{Experimental Setup}
\label{sec:exp_setup}

\noindent\textbf{Hardware configuration.}
All experiments are conducted on a workstation with an Intel Core i9-13900 CPU (\SI{2.0}{\giga\hertz} base, 24 cores), \SI{128}{\giga\byte} DDR5 RAM, and a 1\,TB KIOXIA NVMe SSD,
running Ubuntu 24.04.

\noindent\textbf{Baselines.}
We compare LAANN with five SOTA disk-based ANNS methods: DiskANN~\cite{diskann}, Starling~\cite{starling}, MARGO~\cite{margo}, PipeANN~\cite{pipeann}, and PageANN~\cite{pageann}.

\noindent\textbf{Datasets.}
All schemes are evaluated on five representative benchmark datasets covering diverse scales and vector dimensionalities: SIFT100M~\cite{SIFT}, SPACEV100M~\cite{SPACEV}, DEEP100M~\cite{DEEP}, SPACEV1B~\cite{SPACEV}, and SIFT1B~\cite{SIFT}, as shown in Table~\ref{tab:dataset}.

\noindent\textbf{Evaluation metrics.}
Our evaluation focuses on three primary performance indicators. Recall@10 measures the proportion of ground truth nearest neighbors found within the top-10 retrieved results. We report query latency as the mean processing time per query in milliseconds (ms). System throughput is measured as queries per second (QPS) under concurrent workloads.

\noindent\textbf{Parameters.}
All systems are configured under the same hardware, dataset, and index construction parameters, ensuring identical disk usage and in-memory footprint. Specifically, we use consistent parameters for average vector degree, vector compression ratio, candidate pool size during construction, and number of vectors per page across all methods. During search, the search beamwidth $W$ (i.e., the batch size of I/O read requests) is set to 5 and the number of query threads is fixed at 16 across all datasets. For PipeANN, whose beamwidth is dynamic rather than fixed, $W$ is set to 32 (the maximum value recommended by the authors). For LAANN, the beamwidth of normal search mode is set to 5, while the spike ratio $\alpha = 25\%$, decay ratio $\beta = 0.95$, and candidate enlargement ratio $\mu = 2.4$. For other method-specific hyperparameters, we adopt optimal values recommended by the respective papers or official repositories.

\begin{table}[!t]
\centering
\caption{Characteristics of datasets.}
\resizebox{0.99\columnwidth}{!}{
\begin{tabular}{lcccc}
\hline
Dataset      & \# of vectors  & Dimension & Type & \# of queries  \\
\hline
SIFT100M~\cite{SIFT}     & 100M   & 128 & uint8 & 10000  \\
SPACEV100M~\cite{SPACEV}   & 100M   & 100 & int8 & 29316   \\
DEEP100M~\cite{DEEP} & 100M   & 96 & float & 10000       \\
SIFT1B~\cite{SIFT}     & 1B   & 128 & uint8 & 10000  \\
SPACEV1B~\cite{SPACEV}   & 1B   & 100 & int8  & 29316 \\
\hline
\end{tabular}
}
\label{tab:dataset}
\end{table}

\begin{table*}[!t]
\centering
\small
\caption{Comparison of throughput, latency, and average I/Os at
Recall@10 = 0.9 with a memory usage of 0.5$\times$ dataset size.}
\label{tab:million_scale_results}
\resizebox{1\textwidth}{!}{%
\begin{tabular}{lccccccccc}
\toprule
\multirow{2}{*}{Scheme / Dataset} &
\multicolumn{3}{c}{SIFT100M} &
\multicolumn{3}{c}{SPACEV100M} &
\multicolumn{3}{c}{DEEP100M} \\
\cmidrule(lr){2-4} \cmidrule(lr){5-7} \cmidrule(lr){8-10}
 & Throughput (QPS) & Latency (ms) & Mean I/Os
 & Throughput (QPS) & Latency (ms) & Mean I/Os
 & Throughput (QPS) & Latency (ms) & Mean I/Os \\
\midrule
DiskANN~\cite{diskann}
  & 1101.11 & 14.49 & 153.74
  & 1714.20 & 9.29  & 99.60
  & 2833.08 & 5.59  & 54.96 \\
Starling~\cite{starling}
  & 2172.97 & 7.26  & 74.35
  & 3472.86 & 4.53  & 45.38
  & 3740.92 & 4.21  & 38.69 \\
MARGO~\cite{margo}
  & 2188.71 & 7.20  & 73.66
  & 2975.00 & 5.32  & 48.20
  & 3656.17 & 4.32  & 37.48 \\
PipeANN~\cite{pipeann}
  & 701.04  & 22.71 & 159.53
  & 763.38  & 20.85 & 146.01
  & 1266.06 & 12.55 & 85.53 \\
PageANN~\cite{pageann}
  & 2718.40 & 5.87  & 57.30
  & 3761.60 & 4.24  & 37.50
  & 4857.62 & 3.28  & 27.20 \\
LAANN (ours)
  & \textbf{3825.60} & \textbf{4.17} & \textbf{36.10}
  & \textbf{7987.20} & \textbf{1.99} & \textbf{15.70}
  & \textbf{9516.90} & \textbf{1.67} & \textbf{11.40} \\
\bottomrule
\end{tabular}%
}
\end{table*}

\subsection{Performance on 100-M Scale Datasets}

We evaluate all methods under a memory budget of $0.5\times$ the dataset size. Within this budget, all methods allocate $0.2\times$ for storing compressed vectors and $0.2\times$ for storing cached disk pages. Methods with an in-memory graph index (Starling, MARGO, Page-ANN, and LAANN) allocate an additional $0.1\times$ for the index. DiskANN, which does not maintain an in-memory graph index, allocates the full remaining $0.3\times$ for cached disk pages. PipeANN does not support storing cached disk pages, so its full remaining $0.3\times$ is allocated to its in-memory graph index.

Figure~\ref{fig:overall_performance_100m} shows the recall-latency and recall-QPS tradeoff curves across three 100M-scale datasets. LAANN consistently achieves the best tradeoff across all datasets, delivering significantly lower latency and higher throughput than all baselines at the same recall level. The advantage is particularly pronounced on Spacev100m and Deep100m, where LAANN achieves over $2\times$ higher QPS than the second-best method across a wide range of recall targets. 

Notably, PipeANN consistently underperforms all other methods despite maintaining an in-memory graph index. As shown in Figure~\ref{fig:io_cpu_breakdown} (a), PipeANN achieves competitive latency at $T=2$ threads, but its latency degrades most sharply as thread count increases. The root cause is that PipeANN maximizes I/O parallelism at the cost of issuing more I/O operations, which becomes increasingly expensive under I/O contention at higher thread counts. In contrast, LAANN issues 77\%--89\% fewer I/O operations per query than PipeANN (Table~\ref{tab:million_scale_results}), so each thread places a proportionally smaller load on the SSD, making LAANN inherently robust to I/O contention as thread count increases.

Table~\ref{tab:million_scale_results} provides a detailed comparison at Recall@10 = 0.9. LAANN outperforms the best baseline PageANN by $1.41\times$--$2.12\times$ in throughput across all three datasets, while reducing mean I/Os by $37\%$--$58\%$. Notably, the throughput gain is consistently larger than the I/O reduction alone would suggest, indicating that LAANN benefits not only from fewer I/Os but also from better CPU utilization through the priority I/O-CPU pipeline. Compared to DiskANN, which represents the most widely deployed baseline, LAANN achieves $3.47\times$--$4.66\times$ higher throughput, demonstrating the fundamental advantage of I/O-aware search over standard greedy beam search.

\subsection{Performance on Billion-Scale Datasets}
We evaluate LAANN against DiskANN, PipeANN, and PageANN on SIFT1B and SpaceV1B under a memory budget of $0.3\times$ the dataset size. Starling and MARGO are excluded as their construction or search phase exceeds the memory capacity of our setup. Within this budget, $0.2\times$ is reserved for compressed vectors across all methods. For the remaining $0.1\times$, PageANN and PipeANN allocate it fully to their in-memory graph index, which stores both graph edges and raw vector values of the sampled nodes. LAANN uses only $0.05\times$ for its lightweight in-memory graph index since it searches using compressed vectors and does not need to store raw vector values, freeing the remaining $0.05\times$ for additional cached disk pages. All three in-memory graph indexes are built over the same number of sampled nodes to ensure a fair comparison. DiskANN does not maintain an in-memory graph index and allocates the full remaining $0.1\times$ to cached disk pages.

\begin{figure}[!h]
    \centering
    \includegraphics[width=0.99\linewidth]{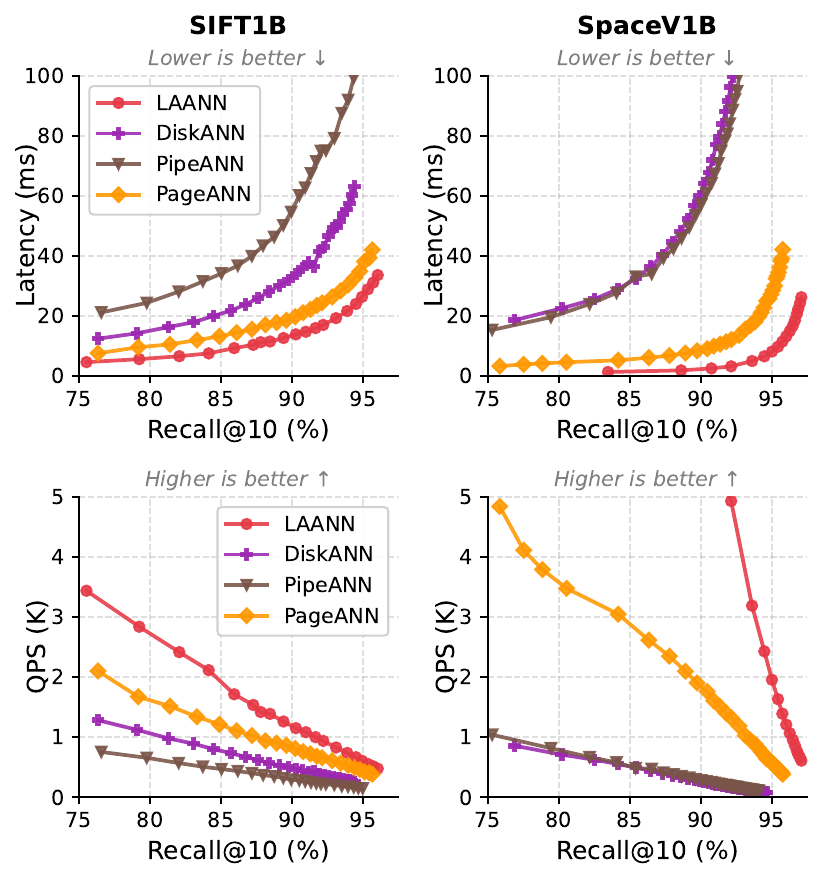}
    \caption{Recall@10 vs. latency (top row) and recall@10
    vs. throughput (bottom row) on SIFT1B and SpaceV1B
    under a memory budget of $0.3\times$ the dataset size.}
    \label{fig:overall_performance_1b}
\end{figure}

As shown in Figure~\ref{fig:overall_performance_1b}, LAANN consistently achieves the lowest latency and the highest throughput across the entire recall range on both datasets. On SIFT1B at Recall@10 = 0.9, LAANN achieves 1,147.7 QPS at 13.87ms, outperforming PageANN by $1.42\times$, DiskANN by $2.38\times$, and PipeANN by $4.03\times$ in throughput. On SpaceV1B, the advantage is even more pronounced. Compared to the 100M-scale results, the performance gap between LAANN and all baselines widens significantly at billion scale. This is because billion-scale datasets produce longer search paths and more I/O operations per query, intensifying I/O contention and increasing per-request I/O latency. Under these conditions, LAANN's ability to reduce the total number of I/O operations translates into a proportionally larger performance advantage.

\subsection{Impact of Cached Graph Nodes}
In this experiment, we evaluate how effectively each method utilizes cached graph nodes to improve search performance. Specifically, we vary the volume of cached graph nodes by adjusting the memory budget allocated to storing them, from $0.3\times$ to $0.9\times$ the dataset size in total. Within each budget, $0.2\times$ is reserved for compressed vectors and, for methods that support an in-memory graph index (Starling, MARGO, PageANN, and LAANN), an additional $0.1\times$ is allocated for the index. The remaining budget determines the number of graph nodes that can be cached in memory. DiskANN, which does not maintain an in-memory graph index, allocates the full remaining budget to cached graph nodes. PipeANN is excluded from this experiment as it does not support caching graph nodes to accelerate the search.

\begin{figure}[!t]
    \centering
    \includegraphics[width=0.99\linewidth]{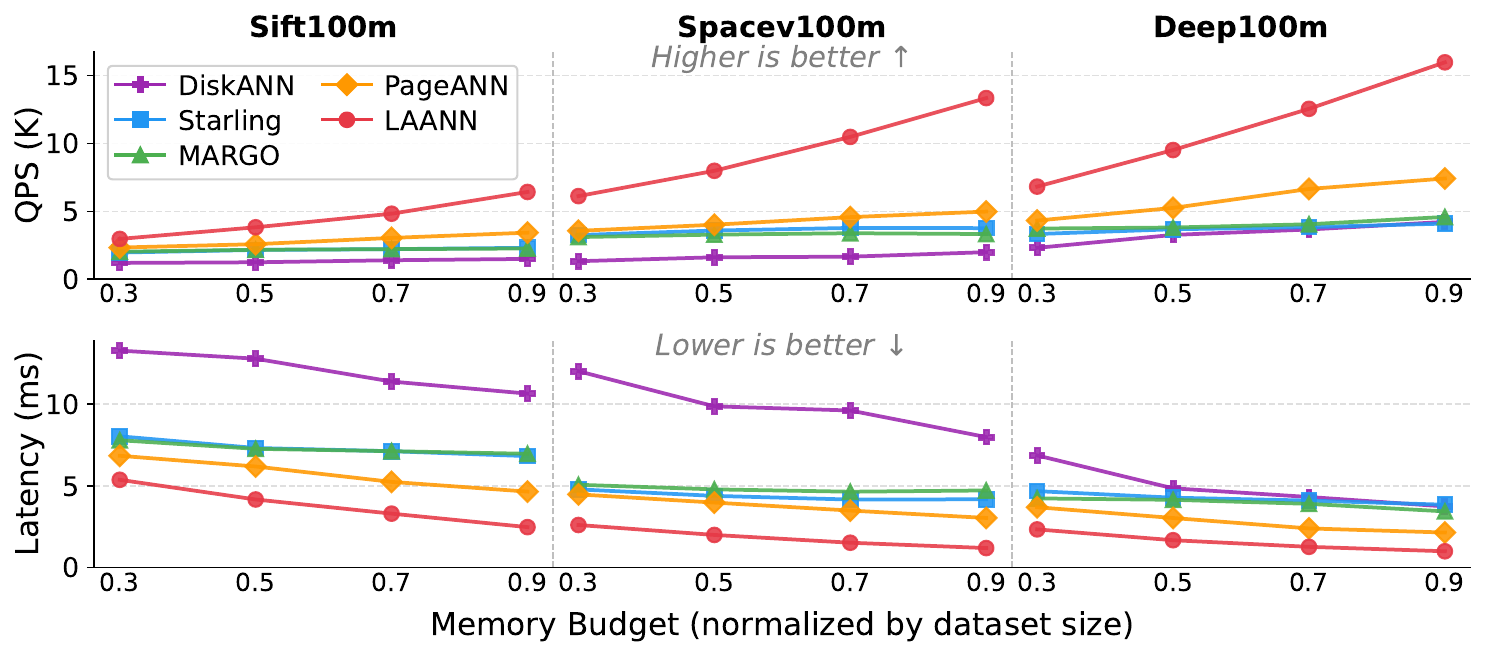}
    \caption{QPS (K) and latency at Recall@10 = 0.9 under varying cached graph node budgets on three 100M-scale datasets. PipeANN is excluded as it does not support caching graph nodes.}
    \label{fig:memory_budget_impact}
\end{figure}

As shown in Figure~\ref{fig:memory_budget_impact}, LAANN consistently outperforms all baselines in both QPS and latency across all three datasets and memory budgets. More importantly, LAANN's performance improves significantly as the memory budget increases, while the baselines, particularly DiskANN, Starling, and MARGO, show only marginal improvement. On Spacev100m, for instance, LAANN's QPS increases from 6,122.6 at $0.3\times$ to 13,346.6 at $0.9\times$, a $2.18\times$ improvement, while Starling improves by only $1.16\times$ over the same range. This stark contrast demonstrates that LAANN makes significantly better use of additional cached data than existing methods. The reason is fundamental: existing methods expand nodes in strict distance order regardless of whether they are cached, so additional cached data does not change the search path. LAANN, by contrast, actively exploits cached neighbors through the look-ahead search and priority IO-CPU pipeline, directly translating additional memory into fewer I/Os and lower latency.

\subsection{Breakdown Analysis}
We analyze the individual contribution of each LAANN component by starting from PageANN as the baseline and adding components one at a time. The baseline PageANN caches the same number of disk pages as LAANN, but uses \textbf{no in-memory graph index} and the standard greedy beam search algorithm without any of LAANN's optimizations. Note that PageANN in Table~\ref{tab:million_scale_results} benefits from its own in-memory graph index, which is the primary reason its QPS there (2718.4 on SIFT100M; 3761.6 on SPACEV100M) is higher than the baseline here (1825.8 on SIFT100M; 2686.6 on SPACEV100M). This controlled setup ensures that each row in Table~\ref{tab:breakdown} isolates exactly one algorithmic contribution. The percentage shown in each cell reflects the individual contribution of that component relative to the baseline.

\begin{table}[!t]
\centering
\small
\caption{Breakdown analysis: Individual contribution of each LAANN component at Recall@10 = 0.9 on SIFT100M and SPACEV100M.}
\label{tab:breakdown}
\resizebox{\columnwidth}{!}{%
\begin{tabular}{lcccc}
\toprule
\multicolumn{5}{c}{\textbf{SIFT100M}} \\
\midrule
Configuration & QPS & Latency (ms) & I/O Latency (ms) & Mean I/Os \\
\midrule
(a) PageANN (baseline)          & 1825.8                   & 8.74                  & 7.74                  & 66.8 \\
(b) + Look-ahead search         & \textbf{3097.1 (+70\%)}  & \textbf{5.14 (-41\%)} & \textbf{3.84 (-50\%)} & \textbf{46.4 (-31\%)} \\
(c) + Priority I/O-CPU pipeline & 3347.5 (+14\%)           & 4.76 (-4\%)           & 3.00 (-11\%)          & 40.6 (-9\%) \\
(d) + Lightweight in-memory index   & 3889.2 (+30\%)           & 4.09 (-8\%)           & 2.82 (-15\%)          & 36.1 (-13\%) \\
\midrule
\multicolumn{5}{c}{\textbf{SPACEV100M}} \\
\midrule
Configuration & QPS & Latency (ms) & I/O Latency (ms) & Mean I/Os \\
\midrule
(a) PageANN (baseline)          & 2686.6                   & 5.94                  & 4.81                  & 45.2 \\
(b) + Look-ahead search         & \textbf{5267.2 (+96\%)}  & \textbf{3.03 (-49\%)} & \textbf{2.04 (-58\%)} & \textbf{25.8 (-43\%)} \\
(c) + Priority I/O-CPU pipeline & 6090.8 (+30\%)           & 2.62 (-7\%)           & 1.35 (-14\%)          & 20.6 (-12\%) \\
(d) + Lightweight in-memory index   & 8208.5 (+80\%)           & 1.94 (-11\%)          & 1.15 (-4\%)           & 15.7 (-10\%) \\
\bottomrule
\end{tabular}%
}
\end{table}

\noindent\textbf{Look-ahead search.} The look-ahead search contributes the largest individual improvement across all metrics. On SIFT100M, it improves QPS by 70\%, reduces latency by 41\%, and reduces mean I/Os by 31\% relative to the baseline. On SPACEV100M, the gains are even more pronounced, with QPS improving by 96\%, latency reducing by 49\%, and mean I/Os reducing by 43\%. This confirms that the look-ahead search is the most impactful component, as it directly reduces the number of I/O operations by prioritizing in-memory neighbors in the approach phase and maximizing I/O parallelism in the convergence phase.

\noindent\textbf{Priority I/O-CPU pipeline.} The priority I/O-CPU pipeline with the customized candidate pool contributes a further 14\% QPS improvement and 9\% I/O reduction on SIFT100M, and 30\% QPS improvement and 12\% I/O reduction on SPACEV100M. The I/O latency reduction is particularly notable, as the pipeline hides more computation behind I/O waiting time and improves I/O efficiency through better prioritization of CPU work.

\noindent\textbf{Lightweight in-memory graph index.} The lightweight in-memory graph index contributes a further 30\% QPS improvement and 13\% I/O reduction on SIFT100M, and 80\% QPS improvement and 10\% I/O reduction on SPACEV100M. The substantially larger QPS gain on SPACEV100M demonstrates that seeding the disk graph candidate pool with high-quality candidates at the start of the search is particularly effective on datasets where the approach phase dominates, as it significantly reduces the number of approach phase I/Os and hastens convergence.

Overall, all three components contribute meaningfully and complementarily to LAANN's performance. The look-ahead search provides the largest single gain by directly reducing I/Os, the pipeline further improves efficiency by exploiting CPU idle time, and the lightweight in-memory graph index accelerates convergence from the very start of the search.

\section{Related Work}\label{sec:related}

\noindent\textbf{Distributed ANNS approaches.} Distributed ANNS systems scale to web-scale datasets by partitioning the index across multiple nodes. Independent sharding~\cite{milvus, deng2019pyramid, wei2020analyticdb} assigns disjoint data partitions to each node and merges per-shard results, trading query quality for simplicity. Global indexing approaches~\cite{zhi2025towards} build a unified index across nodes to avoid redundant computation at the cost of higher communication overhead. Both categories target inter-machine scalability and are complementary to LAANN, which focuses on maximizing single-machine I/O efficiency. LAANN's I/O-aware search could serve as the per-node search engine in such systems.

\noindent\textbf{Disk-based ANNS approaches.} Disk-based methods offload vectors and graph edges to SSD to reduce memory requirements, as discussed in Section~\ref{sec:background}. DiskANN~\cite{diskann} introduced the foundational design of storing compressed vectors in memory and fetching full-precision vectors from disk on demand, with caching of frequently visited nodes to reduce I/Os. Starling~\cite{starling} and MARGO~\cite{margo} improve spatial locality within each disk page and maintain in-memory entry point graphs to reduce hops to the nearest-neighbor region. PageANN~\cite{pageann} eliminates I/O read amplification by aligning graph node granularity with SSD page size. The most closely related work is PipeANN~\cite{pipeann}, which also distinguishes between search phases and adjusts the beam width in the convergence phase. However, PipeANN minimizes CPU work before issuing I/O in order to maximize I/O overlap, at the cost of issuing more I/O operations, which causes severe latency degradation under I/O contention. In contrast, LAANN prioritizes processing relevant CPU work before making I/O decisions, both on cached vectors and on vectors just loaded from disk, in order to reduce the total number of I/O operations. Furthermore, PipeANN increases beam width linearly in the convergence phase without accounting for the increasing eviction probability of vectors near the end of the candidate pool, causing I/O waste near the end and I/O submission delays at the start. LAANN's dynamic beam width scheme addresses this by spiking aggressively at the start and decaying gradually. Fundamentally, all existing methods, including PipeANN, optimize I/O from the I/O perspective alone and expand nodes in strict distance order regardless of whether data is cached. LAANN is the first to exploit the coupling between CPU computation and I/O operations, achieving a better CPU-I/O balance and lower overall query latency.

\noindent\textbf{Specialized hardware ANNS approaches.} Several systems co-design search algorithms with specialized hardware to bypass CPU-I/O bottlenecks. FusionANNS~\cite{tian2025towards} offloads distance computation to GPUs to overlap with SSD I/O. HM-ANN~\cite{hm-ann} places upper graph layers in persistent memory and lower layers in DRAM to exploit the memory capacity hierarchy. SmartANN~\cite{smartann} pushes computation directly into the storage device to eliminate host I/O entirely. While these approaches achieve strong results on their target hardware, they are inherently tied to specific hardware configurations. LAANN targets the same I/O bottleneck through purely algorithmic means, remaining hardware-agnostic and deployable on any commodity server.

\section{Conclusion}\label{sec:conclusion}
We presented LAANN, a disk-based ANNS system that makes the search algorithm I/O-aware by identifying the tight coupling between CPU work and I/O operations and the fundamentally different I/O compositions across search phases. LAANN introduces three complementary techniques: a look-ahead search that adapts its strategy to the current search phase, a priority I/O-CPU pipeline that fills CPU idle time with work ordered by relevance to the next I/O decision supported by a customized candidate pool with an overflow area, and a fast lightweight in-memory graph index that seeds the disk graph search with high-quality candidates from the start. Experiments on three 100M-scale and two billion-scale datasets demonstrate that LAANN achieves $1.41\times$--$4.66\times$ higher throughput, 29\%--79\% lower latency, and $1.59\times$--$6.34\times$ fewer I/O operations compared to SOTA baselines, confirming that making the search algorithm I/O-aware is a fundamental improvement over distance-ordered greedy search for disk-based ANNS.

\section{Acknowledgment}
This work was partially supported by NSF 2343863, 2413520, 2417747 and 2440611. Any opinions, conclusions, or recommendations expressed in this material are those of the authors and do not necessarily reflect the views of the NSF.

\bibliographystyle{ACM-Reference-Format}
\bibliography{ref}

\end{document}